\documentclass[preprint]{article}
\usepackage{spconf}
\usepackage{amsmath, graphicx}

\usepackage[hidelinks]{hyperref}
\usepackage{csquotes}
\usepackage{cleveref}

\usepackage{booktabs}
\usepackage{tabularx}

\usepackage{multirow}

\usepackage{xcolor}

\usepackage[acronym]{glossaries} 

\newacronym{dnn}{DNN}{Deep Neural Network}
\newacronym{mlp}{MLP}{Multilayer Perceptron}
\newacronym{cnn}{CNN}{Convolutional Neural Network}
\newacronym{rnn}{RNN}{Recurrent Neural Network}
\newacronym{crnn}{CRNN}{Convolutional Recurrent Neural Network}
\newacronym{asr}{ASR}{Automatic Speech Recognition}
\newacronym{ema}{EMA}{Exponential Moving Average}
\newacronym{lstm}{LSTM}{Long Short-term Memory}
\newacronym{apc}{APC}{Autoregressive Predictive Coding}

\usepackage{siunitx}						
\DeclareSIUnit\byte{B}

\title{Self-supervised Pretraining for Robust Personalized Voice Activity Detection in Adverse Conditions}

\copyrightnotice{\copyright\ IEEE 2024}
\toappear{To appear in {\it Proc.\ ICASSP2024, April 14th, 2024, Seoul, South Korea}}

\name{Holger Severin Bovbjerg$^{1}$, Jesper Jensen$^{1, 2}$, Jan Østergaard\,$^{1}$, Zheng-Hua Tan$^{1, 3}$} 
\address{$^{1}$Department of Electronic Systems, Aalborg University, Denmark\\
$^{2}$Oticon A/S, Denmark\\
$^{3}$Pioneer Centre for AI, Denmark\\
\texttt{\{hsbo,jje,jo,zt\}@es.aau.dk}}
\begin{document}
\ninept
\maketitle
\begin{abstract}
In this paper, we propose the use of self-supervised pretraining on a large unlabelled data set to improve the performance of a personalized voice activity detection (VAD) model in adverse conditions.
We pretrain a long short-term memory (LSTM)-encoder using the autoregressive predictive coding (APC) framework and fine-tune it for personalized VAD.
We also propose a denoising variant of APC, with the goal of improving the robustness of personalized VAD.
The trained models are systematically evaluated on both clean speech and speech contaminated by various types of noise at different SNR-levels and compared to a purely supervised model.
Our experiments show that self-supervised pretraining not only improves performance in clean conditions, but also yields models which are more robust to adverse conditions compared to purely supervised learning. 
\end{abstract}
\section{Introduction}
\label{sec:intro}
Being able to detect the presence of speech in a potentially noisy signal is a commonly utilized processing step in modern speech processing systems, generally referred to as voice activity detection (VAD). 
A VAD system has to classify whether a short frame of audio contains speech, usually from speech features like Mel-filterbank features, in an unsupervised \cite{TAN_rVAD} or supervised manner \cite{dinkel2021voice}. 
Applications include using a VAD model as a preprocessing step for automatic speech recognition, as a gating mechanism for microphones in online meeting devices, or as a part of a speech enhancement system. 
For real-time applications, such as speech enhancement for hearing aids, it is desired that the VAD model is low-latency and low complexity, while also being robust in adverse conditions such as background noise. 

Recently, personalized VAD models, which are also able to determine whether the speech is from a target speaker, have been proposed \cite{ding20_PVAD1, Medennikov_TS_VAD, Maokui_TS_VAD_improved_i_vector, ding22_PVAD2}. 
These personalized VAD models introduce a number of interesting capabilities, such as removing false-triggering on background speakers, at the expense of also needing to model speaker characteristics.
Training a model to detect voice activity and distinguish between target speech and non-target speech in a supervised fashion requires a large amount of speech data from several speakers annotated with both VAD labels and framewise speaker identity.
For example, \cite{ding20_PVAD1} and \cite{Medennikov_TS_VAD} utilize \SI{960}{\hour} of annotated speech from \num{2338} different speakers to train their models, while \cite{ding22_PVAD2} utilize up to \SI{27500}{\hour} of annotated speech.
Although high-quality VAD labels can be automatically obtained using forced-alignment \cite{kraljevski_forced_alignment_vs_human_vad}, relatively clean speech and a corresponding transcript of the speech signal is required. 
Additionally, frame-level speaker identity labels can be difficult to obtain.
Therefore, the adoption of Personalized VAD models is limited by the ability to obtain such large labelled data sets, restricting their widespread adoption.

Self-supervised learning (SSL) methods provide a means of utilizing unlabelled data, which is easier to obtain in large quantities.  
Models pretrained using SSL have shown state-of-the-art performance in many domains, including speech processing \cite{baevski_wav2vec2, HuBERT_paper, Chen_WavLMLS}, and have also been shown to learn more robust features than purely supervised models \cite{liu_robust_selfsupervised, luo_understanding_SSL_robustness}. 
The application of SSL for pretraining of VAD models is currently unexplored, although one study used speech features from a pretrained wav2vec2 \cite{baevski_wav2vec2} model as input features to a VAD model, and found that they perform better than standard Mel-spectrogram features \cite{Alisamir_CrossdomainVA}.
However, using a large pretrained speech model (95M+ parameters) for feature extraction arguably defeats the purpose of having a small, efficient VAD model.

In this work, we propose to use a simple SSL framework known as Autoregressive Predictive Coding (APC), to directly pretrain a small LSTM-encoder, with the aim of improving performance and robustness in adverse conditions, when fine-tuning for personalized VAD.
Additionally, we propose a denoising variant of APC for improved robustness.
Here, we modify the APC framework to predict future clean speech frames from noisy input features, as opposed to predicting clean future frames from clean input features.

We carry out experiments, using both clean and noisy training data generated through online multistyle training (MTR) \cite{Prabhavalkar_multistyle_training}. 
The trained models are evaluated systematically on both clean test data and on different test sets containing either seen or unseen noise at SNR levels ranging from \SIrange{-5}{20}{\decibel}. 
The results show the following:
\begin{enumerate}
    \setlength\itemsep{0.2em}
    \item APC pretraining and fine-tuning on clean data, leads to an absolute improvement in mean average precision (mAP) of \SI{1.9}{\percent} compared to supervised training, when evaluating the models in clean conditions. 
    Interestingly, an average absolute improvement of \SI{6.05}{\percent} is observed for noisy conditions, while not having seen any noise during training. 
    \item When using MTR, APC pretraining leads to an average absolute improvement for seen noise \SI{4.8}{\percent} over the baseline, while a further absolute improvement \SI{2.3}{\percent} is observed for the proposed DenoisingAPC. For unseen noise, DenoisingAPC+MTR also achieves the best performance, with an absolute improvement of \SI{8}{\percent} compared to baseline+MTR.
\end{enumerate}

\newpage

The source code used to produce the results of this paper is made publicly available. 
\footnote{\url{https://github.com/HolgerBovbjerg/SelfSupervisedPVAD}}

\section{Methodology and Data Sets}
This section describes our personalized VAD model, the APC pretraining framework and the data used for training and testing.

\subsection{Personalized VAD model}
\begin{figure}[tb]
    \centering
    \includegraphics[scale=0.6]{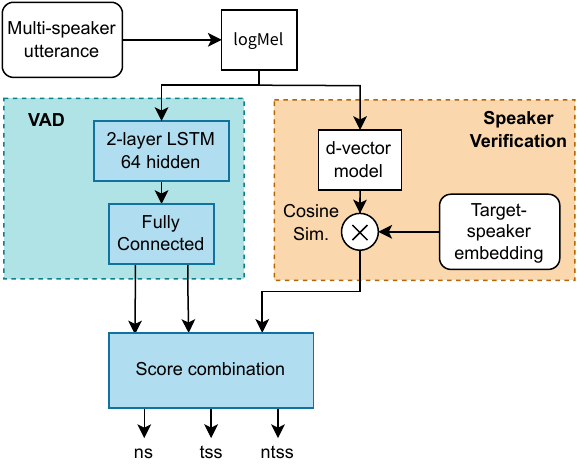}
    \caption{Overview of the personalized VAD system model used in this work. We only train the blue boxes, while we use an already trained d-vector model with fixed weights.}
    \label{fig:PVAD_overview}
\end{figure}

Our personalized VAD model is inspired by the Personal VAD system presented in \cite{ding20_PVAD1} and is illustrated in \Cref{fig:PVAD_overview}.
The personalized VAD classifies input log Mel-filterbank features as either non-speech (ns), target-speaker speech (tss) or non-target-speaker speech (ntss).
We choose to separate the speaker verification and VAD tasks into separate modules and reside to using an already trained model for speaker verification. 
More specifically, we focus on training a robust VAD model, and use a separate already trained d-vector model \cite{Wan_GE2E} to extract speaker embeddings for speaker verification.

The VAD module predicts the probability of speech or no speech, $z^\mathrm{s}$ and $z^\mathrm{ns}$, for any given input frame.
The d-vector model generates an embedding which is compared to the target-speaker embedding through cosine similarity to generate a target-speaker similarity score $s$. 
As $s$ is a similarity score, its value might not necessarily represent the probability of a target-speaker being present.
Therefore, $s$ is scaled by learnable parameters $\alpha$ and $\beta$ such that the scaled similarity becomes $s^\prime = s\alpha + \beta$.
Finally, the VAD output and scaled similarity score are combined such that
\begin{equation}
    z^k_t =
    \begin{cases}
        z^\mathrm{ns}_t & k = \mathrm{ns}, \\ 
        s^\prime z^\mathrm{s}_t & k = \mathrm{tss}, \\ 
        (1 - s^\prime)z^\mathrm{s}_t & k = \mathrm{ntss}. 
    \end{cases}
\end{equation}
where $z^k_t$ is the output corresponding to class $k$ at time frame $t$ and is used to classify each frame as either non-speech (ns), target-speaker speech (tss) or non-target-speaker speech (ntss).

In \cite{ding20_PVAD1}, the authors also propose a personalized VAD where the target-speaker embedding is simply concatenated to the features, which is then used as input to the VAD.
Here, the VAD model also learns to extract speaker characteristics through implicit knowledge distillation from the speaker embedding model, removing the need for a separate speaker embedding model during runtime. 
However, our implementation failed to achieve good performance using this approach.

For the speaker embedding model, we use a freely available d-vector model as described in \cite{Wan_GE2E}. 
The d-vector model used in this work, is pretrained on VoxCeleb \cite{NAGRANI_voxceleb} and LibriSpeech-other data. 
It has 3 LSTM layers, each with a hidden dimension of 256 producing 256-dimensional d-vector speaker embeddings and has a total of 1.4M parameters. 
As in \cite{ding20_PVAD1}, our VAD model is a 2-layer LSTM with a hidden dimension of 64, yielding a total of 60k parameters. 
Both the d-vector and VAD models take 40-dimensional log Mel-filterbank features as input, computed from \SI{25}{\milli\second} frames with frame shift of \SI{10}{\milli\second}. 

Target-speakers are enrolled by generating a d-vector embedding using one or more enrolment utterances (minimum \SI{5}{\second}). 
Here, speaker embeddings are computed by sliding a \SI{1.6}{\second} window across the enrolment utterances with a shift of \SI{0.4}{\second}, generating an embedding for each window position, which are then averaged to generate the target-speaker d-vector embedding.


\subsection{Autoregressive predictive coding}
\begin{figure}[tb]
    \centering
    \includegraphics[scale=0.6]{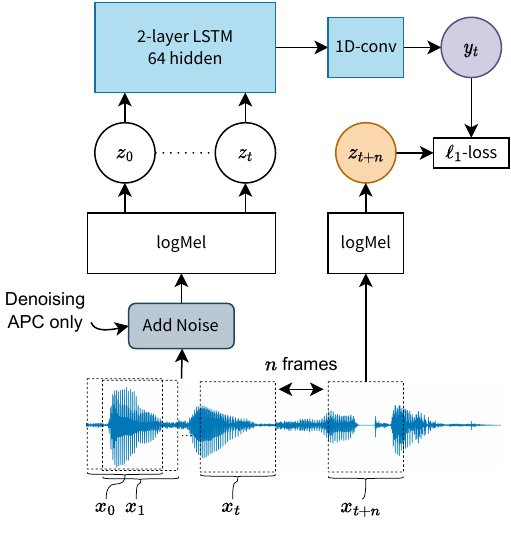}
    \caption{Overview of the APC framework and DenoisingAPC variant. Note that noise is only added for DenoisingAPC, while this is skipped for standard APC.}
    \label{fig:APC_overview}
\end{figure}

Inspired by the success of pretraining language models, the Autoregressive Predictive Coding \cite{unsupervised_AR_model_for_speech} framework predicts future speech features from the current and previous feature vectors.
More formally, given a sequence of feature vectors $z_0, \dots, z_t$ computed from frames $x_0, \dots, x_t$, we ideally seek a model $f$ that predicts $z_{t+n}$ such that $y_t = f(z_0, \dots, z_t) = z_{t + n}$, denoting $z_t$ as the feature vector and $y_t$ as the output at time frame $t$.
APC thus builds on the notion that if a model is able to predict the future, it must have a good representation of the past and present.
An overview of the APC framework, including the denoising variant, is illustrated in \Cref{fig:APC_overview}.
Models pretrained using APC have been shown to learn both speaker and content information and shown good performance for a number of downstream tasks \cite{APC_paper}.

While more complex SSL methods such as wav2vec2 \cite{baevski_wav2vec2} can also learn information from future frames, APC only encodes information from previous frames. 
Thus, the learned representation does not rely on future information.
This is particularly suitable for VAD applied to real-time applications, as the VAD model is desired to be causal.
In our experiments, we feed the log Mel-filterbank features to a 2-layer LSTM encoder, and use a single 1D-convolutional layer to project the hidden representations back to the input feature space.

In addition to the standard APC framework, we also propose a denoising variant of APC. 
Here, speech features are extracted from both clean speech and speech which have been corrupted by noise.
We then predict future clean features from noisy input features, as depicted in \Cref{fig:APC_overview}.
This forces the model to extract information related to the source signal from a noisy mixture, thus learning to distinguish the source signal from background noise.

\subsection{Data sets}
As mentioned in \cite{ding20_PVAD1}, the amount of readily available multi-speaker data with natural speaker turns, as well as speaker identity information is limited, and as a result we carry out experiments on a simulated multi-speaker data set.
While speakers might overlap in realistic settings, such as a cocktail party scenario, it has been found that a personalized VAD model trained on non-overlapping speech also performs well on overlapping speech \cite{Medennikov_TS_VAD}.
Following \cite{ding20_PVAD1}, we uniformly sample 1 to 3 utterances from individual speakers and a target-speaker is randomly selected from one of the individual utterances.
We then simply concatenate the utterances to generate multi-speaker utterances. 

For our experiments, we use the freely available Librispeech \cite{librispeech} data set to construct both training and test data.
Librispeech consists of \SI{960}{\hour} training data split into two sets of \SI{100}{\hour} and \SI{360}{\hour} categorized as \enquote{clean}, and a \SI{500}{\hour} \enquote{other} set, which is \enquote{less clean}.
Similarly, both a \enquote{clean} and an \enquote{other} set is available for testing.
\Cref{tab:data_summary} shows a summary of the different data sets used in our experiments for training, pretraining and testing.

In our experiments, we only use utterances within the same set when constructing multi-speaker utterances. 
For pretraining, the multi-speaker utterances constructed from the \textit{train-clean-100}, \textit{train-clean-360} and \textit{train-other-500} sets are used, yielding a total of all \SI{960}{\hour} speech. 
For supervised training, only multi-speaker utterances generated from the \SI{100}{\hour} \textit{train-clean-100} set are used.
Additionally, we also train the models on the 10h LibriLight \cite{librilight} training set.
This simulates a setting with a large pool of unannotated data available for pretraining and a smaller pool of labelled data available for supervised training.
As Librispeech includes speech transcripts, VAD labels are generated using forced-alignment \cite{montreal_forced_aligner}, while we generate framewise speaker labels using the speaker identity information included in the Librispeech metadata.

\begin{table}[b]
\centering
\caption{Summary of data sets with concatenated utterances used for training, pretraining and testing.}
\label{tab:data_summary}
\footnotesize
\begin{tabular}{@{}llll@{}}
\toprule
Set           & \#Multi-speaker Utt.    & \#Speakers & Used for \\ \midrule
train-100-clean & 14252              & 251  & SSL + Supervised \\
train-360-clean & 52068              & 921  & SSL \\
train-500-other & 74177              & 1166 & SSL \\
LibriLight 10h & 1364              & 44 & Supervised \\
test-clean      & 1315              & 40   & Testing \\ \bottomrule
\end{tabular}
\end{table}

For testing the model performance in clean conditions, we use the utterances generated from the \textit{test-clean} set.
To be able to evaluate the trained model in varying adverse conditions, noisy test data is generated by adding noise, to the \textit{test-clean} multi-speaker utterances. 
Here, we pick two environmental noise types, namely bus and café, and two speech-like noise types, namely babble and speech-shaped noise, each representing a realistic adverse condition.
For each noise type, noisy tests set have been generated by adding the noise type at a specific SNR level, ranging from \SIrange{-5}{20}{\decibel} in steps of \SI{5}{\decibel}, yielding a total of 24 noisy test sets. 

\subsection{Supervised baseline}
When training the personalized VAD model, the pretrained d-vector model weights are fixed, and we only update the VAD network and fully-connected network depicted in \Cref{fig:PVAD_overview}. 
We reuse the hyperparameter choices in \cite{ding20_PVAD1} using a cross-entropy loss and a batch size of 64, the ADAM optimizer with an initial learning rate of $5\cdot10^{-5}$, and a gradual reduction of the learning rate following a cosine annealing schedule. 

\subsection{Pretraining and fine-tuning}
During pretraining, only the LSTM-encoder in the VAD network depicted in \Cref{fig:PVAD_overview} is pretrained, yielding a system as seen in \Cref{fig:APC_overview}. 
In \cite{generative_pretraining_for_speech_with_APC} it is found that predicting $n=3$ frames ahead during APC pretraining leads to good downstream task performance, thus we adopt this choice and use an $\ell_1$-loss as the objective function.
We pretrain the LSTM encoder for 10 epochs using a batch size of 32 and ADAM optimizer with an initial learning-rate of 0.01, and a cosine annealing learning rate schedule. 
After pretraining, the LSTM encoder weights are copied to a Personal VAD model which is then fine-tuned, using the same procedure as used for training the supervised baseline.

\subsection{Multistyle training}
A commonly used technique to improve model robustness is adding various noise to the training data.
This technique is generally referred to as multistyle training (MTR) and has been shown to improve model robustness \cite{Prabhavalkar_multistyle_training}.
Therefore, we also carry out experiments where we apply online MTR.
Here, we add noise from different adverse conditions as described in \cite{Kolbaek_noise_robust_speaker_verification}, namely babble, bus, pedestrian, street and speech-shaped noise.
We include babble, bus and speech-shaped noise in both training data and test sets, while keeping café noise unseen during training.
The noise is added at varying SNR levels in the range \SIrange{-5}{20}{\decibel}.
We also add room acoustics from recorded RIRs as used in \cite{Ko_RIR_data_augmentation}. 
When applying MTR, randomly sampled noise and room acoustics are added to the individual multi-speaker utterances, each with a probability of \SI{50}{\percent}.

\subsection{Metrics}
To evaluate the performance of the trained model, we follow \cite{ding20_PVAD1} and compute the average precision score for each class and use the mean average precision (mAP) score as our main evaluation metric. 
For a given class, average precision is computed as
\begin{equation}
    \mathrm{AP}=\sum_n(R_n-R_{n-1})\cdot P_n
\end{equation}
with $P_n$ and $R_n$ being the precision and recall at threshold $n$.
To compute the mAP score, we compute the AP score for each class and take the mean.

\section{Results}
In the following section, the results from our experiments are presented. 
First, we analyse how the trained models perform in clean conditions, followed by an in-depth analysis of how the trained models perform in various adverse environments. All models have been trained using five different random seeds.

\subsection{Clean conditions}
In \Cref{tab:clean_results} the performance of the various models on the clean test set is presented.
Here, ns, tss and ntss denotes no-speech, target-speaker speech and non-targets-speaker speech, respectively.

Comparing the models without MTR, Baseline and APC, the APC pretrained model shows an improvement in mAP of \SI{2.1}{\percent} and DN-APC an improvement of \SI{2.5}{\percent}. 
The DN-APC model scores highest of all models with an mAP of \SI{92.9}{\percent}. 
Using MTR for model robustness usually comes at the cost of a performance drop in clean conditions \cite{Prabhavalkar_multistyle_training}. 
As expected, we observe that the models using MTR perform slightly worse, although the DN-APC model performs comparable to the baseline without MTR. 
When using MTR, the DN-APC pretrained model performs best, with an overall improvement of \SI{1.3}{\percent} compared to the supervised baseline using MTR. 

\begin{table}[t]
\centering
\caption{Results on clean test set (mAP, \%) and \SI{95}{\percent} confidence intervals. Baseline is a purely supervised model, while APC and DN-APC are SSL pretrained models which have been fine-tuned for personalized VAD. MTR denotes that multistyle training was used during supervised training.}
\label{tab:clean_results}
\footnotesize
\begin{tabular}{p{0.15\linewidth}llll}
    \toprule
                                & \multicolumn{3}{c}{AP}   &        \\ 
                                \cline{2-4}
    Model                       & \multicolumn{1}{c}{ns}   & \multicolumn{1}{c}{tss}    & \multicolumn{1}{c}{ntss}   & \multicolumn{1}{c}{mAP}    \\
    \midrule
    Baseline               & $82.5 {\scriptstyle\pm 1.13}$ & $94.5 {\scriptstyle\pm 0.22}$ & $94.8 {\scriptstyle\pm 0.22}$ & $90.6 {\scriptstyle\pm 0.47}$ \\
    +MTR                    & $79.8 {\scriptstyle\pm 0.75}$ & $94.1 {\scriptstyle\pm 0.19}$ & $93.6 {\scriptstyle\pm 0.31}$ & $89.2 {\scriptstyle\pm 0.28}$ \\
    \midrule
    APC                         & $86.9 {\scriptstyle\pm 0.54}$ & $94.9 {\scriptstyle\pm 0.11} $& $95.6 {\scriptstyle\pm 0.01}$ & $92.5 {\scriptstyle\pm 0.23}$ \\
    +MTR                   & $81.9 {\scriptstyle\pm 0.88}$ & $94.4 {\scriptstyle\pm 0.09}$ & $94.2 {\scriptstyle\pm 0.38}$ & $90.1 {\scriptstyle\pm 0.30}$ \\
    \midrule
    DN-APC & $88.0 {\scriptstyle\pm 0.57}$ & $94.7 {\scriptstyle\pm 0.16}$ & $96.1 {\scriptstyle\pm 0.08}$ & $92.9 {\scriptstyle\pm 0.14}$\\
    +MTR                & $81.9 {\scriptstyle\pm 0.31}$ & $94.2 {\scriptstyle\pm 0.30}$ & $95.2 {\scriptstyle\pm 0.30}$ & $90.4 {\scriptstyle\pm 0.10}$ \\
    \bottomrule
\end{tabular}
\end{table}

\subsection{Adverse conditions}
We evaluate the performance of the trained models in various adverse conditions, including background noise consisting of bus, babble, and speech-shaped noise.
Additionally, we also evaluate the performance on a noise type (café noise) unseen during training, to evaluate whether the robustness of the models generalize to an unseen noise type.

\Cref{tab:seen_noise} presents the mAP scores and \SI{95}{\percent} confidence intervals of the trained models when testing for seen noise at different SNR-levels are reported. 
While \Cref{tab:seen_noise} shows summary scores averaged over all noise types, the general picture for each individual noise type is the same. 

Looking \Cref{tab:seen_noise}, the pretrained models clearly outperform the supervised baseline models in seen noise.
Interestingly, the APC model outperforms the baseline by \SI{5.6}{\percent} on average when neither is using MTR, without having seen any noise during pretraining or supervised training. 
Using MTR leads to a further improvement, while DN-APC+MTR yields the best results, outperforming the baseline+MTR by \SI{7.1}{\percent} on average and APC+MTR pretraining by \SI{2.3}{\percent}.   
For \Cref{tab:unseen_noise}, showing results when evaluating the models in unseen noise, a similar pattern is observed, with DN-APC+MTR showing the best results, outperforming the baseline+MTR by \SI{8}{\percent} on average and APC+MTR by \SI{3}{\percent}.  

\begin{table}[tb]
\centering
\caption{Performance in seen noise (mAP, \%) and \SI{95}{\percent} confidence intervals.}
\label{tab:seen_noise}
\footnotesize
\begin{tabular}{@{}lp{0.13\linewidth}p{0.13\linewidth}p{0.13\linewidth}p{0.13\linewidth}p{0.13\linewidth}@{}}
    \toprule
    SNR  & Baseline                                        & Baseline + MTR  & APC                                                                & APC \hphantom{1em} +MTR                               & DN-APC + MTR                          \\ 
    \midrule
    -5   & $63.0 {\scriptstyle\pm 1.51}$               & $67.1 {\scriptstyle\pm 0.81}$               & $65.9 {\scriptstyle\pm 2.22}$            & $72.3 {\scriptstyle\pm 1.48}$             & $74.1 {\scriptstyle\pm 0.99}$                           \\
    0    & $63.8 {\scriptstyle\pm 2.32}$               & $71.0 {\scriptstyle\pm 1.25}$          & $67.3 {\scriptstyle\pm 2.99}$           & $76.9 {\scriptstyle\pm 1.62}$             & $80.0 {\scriptstyle\pm 0.52}$                           \\
    5    & $64.9 {\scriptstyle\pm 2.44}$           & $75.2 {\scriptstyle\pm 1.48}$               & $69.0 {\scriptstyle\pm 2.89}$            & $80.9 {\scriptstyle\pm 0.89}$             & $84.3 {\scriptstyle\pm 0.43}$                           \\
    10   & $66.3 {\scriptstyle\pm 1.72}$           & $79.1 {\scriptstyle\pm 1.73}$               & $71.8 {\scriptstyle\pm 1.78}$            & $84.1 {\scriptstyle\pm 0.44}$             & $86.9 {\scriptstyle\pm 0.35}$                           \\
    15   & $68.7 {\scriptstyle\pm 1.60}$           & $82.5 {\scriptstyle\pm 1.23}$               & $76.6 {\scriptstyle\pm 2.79}$            & $86.5 {\scriptstyle\pm 0.43}$             & $88.2 {\scriptstyle\pm 0.29}$                           \\
    20   & $73.1 {\scriptstyle\pm 1.67}$           & $84.9 {\scriptstyle\pm 1.19}$               & $82.4 {\scriptstyle\pm 3.24}$            & $88.0 {\scriptstyle\pm 0.47}$             & $89.1 {\scriptstyle\pm 0.32}$                           \\
    Avg. & $66.6 {\scriptstyle\pm 1.73}$           & $76.6 {\scriptstyle\pm 1.13}$                & $72.2 {\scriptstyle\pm 1.43}$           & $81.4 {\scriptstyle\pm 0.75}$             & $83.7 {\scriptstyle\pm 0.46}$           \\ 
    \bottomrule
\end{tabular}
\end{table}

\begin{table}[tb]
\centering
\caption{Performance in unseen noise (mAP, \%) and \SI{95}{\percent} confidence intervals.}
\label{tab:unseen_noise}
\footnotesize
\begin{tabular}{@{}lp{0.13\linewidth}p{0.13\linewidth}p{0.13\linewidth}p{0.13\linewidth}p{0.13\linewidth}@{}}
    \toprule
    SNR  & Baseline                                        & Baseline +MTR  & APC                 & APC \hphantom{1em} +MTR                                & DN-APC +MTR                          \\ 
    \midrule
    -5   & $63.3 {\scriptstyle\pm 1.05}$               & $63.7 {\scriptstyle\pm 0.91}$            & $66.0 {\scriptstyle\pm 1.21}$           & $68.3 {\scriptstyle\pm 1.92}$             & $72.8 {\scriptstyle\pm 0.46}$                           \\
    0    & $64.0 {\scriptstyle\pm 1.44}$               & $68.1 {\scriptstyle\pm 1.20}$            & $67.9 {\scriptstyle\pm 1.05}$           & $74.5 {\scriptstyle\pm 2.24}$             & $79.5 {\scriptstyle\pm 0.40}$                           \\
    5    & $65.2 {\scriptstyle\pm 1.56}$               & $73.5 {\scriptstyle\pm 1.14}$            & $70.8 {\scriptstyle\pm 1.33}$           & $80.2 {\scriptstyle\pm 1.69}$             & $84.0 {\scriptstyle\pm 0.31}$                           \\
    10   & $67.3 {\scriptstyle\pm 1.36}$               & $78.7 {\scriptstyle\pm 0.96}$            & $75.3 {\scriptstyle\pm 2.42}$           & $84.3 {\scriptstyle\pm 0.91}$             & $86.6 {\scriptstyle\pm 0.26}$                           \\
    15   & $71.0 {\scriptstyle\pm 1.82}$               & $82.9 {\scriptstyle\pm 0.88}$            & $80.9 {\scriptstyle\pm 3.14}$           & $86.9 {\scriptstyle\pm 0.55}$             & $88.1 {\scriptstyle\pm 0.27}$                           \\
    20   & $77.3 {\scriptstyle\pm 1.19}$               & $85.7 {\scriptstyle\pm 0.74}$            & $86.1 {\scriptstyle\pm 2.30}$           & $88.3 {\scriptstyle\pm 0.44}$             & $89.1 {\scriptstyle\pm 0.34}$                           \\
    Avg. & $68.0 {\scriptstyle\pm 2.19}$               & $75.4 {\scriptstyle\pm 0.82}$            & $74.5 {\scriptstyle\pm 1.44}$           & $80.4 {\scriptstyle\pm 1.16}$             & $83.4 {\scriptstyle\pm 0.29}$           \\ 
    \bottomrule
\end{tabular}
\end{table}

In \Cref{tab:10h_training} the average performance for models trained on LibriLight 10h training set is presented. Here, we observe that the pretrained models outperform the supervised baselines by an even larger margin, with an average improvement of \SI{24.3}{\percent} for DN-APC+MTR compared to baseline+MTR.

\begin{table}[tb]
\centering
\caption{Avg. performance for models trained on 10h LibriLight training set in clean, seen and unseen noise (mAP, \%) and \SI{95}{\percent} confidence intervals.}
\label{tab:10h_training}
\footnotesize
\begin{tabular}{@{}p{0.08\linewidth}p{0.13\linewidth}p{0.13\linewidth}p{0.13\linewidth}p{0.13\linewidth}p{0.13\linewidth}@{}}
    \toprule
    Test set  & Baseline                                    & Baseline +MTR                            & APC                                     & APC \hphantom{1em} +MTR                   & DN-APC +MTR                                             \\ 
    \midrule
    Clean & $57.2 {\scriptstyle\pm 6.96}$               & $55.6 {\scriptstyle\pm 6.47}$            & $83.8 {\scriptstyle\pm 1.89}$           & $82.5 {\scriptstyle\pm 2.46}$             & $83.2 {\scriptstyle\pm 1.91}$           \\ 
    Seen & $54.9 {\scriptstyle\pm 2.37}$               & $55.0 {\scriptstyle\pm 2.43}$            & $63.7 {\scriptstyle\pm 2.05}$           & $68.9 {\scriptstyle\pm 2.07}$             & $77.9 {\scriptstyle\pm 1.57}$           \\ 
    Unseen & $56.2 {\scriptstyle\pm 2.45}$               & $56.3 {\scriptstyle\pm 2.45}$            & $65.0 {\scriptstyle\pm 2.87}$           & $69.6 {\scriptstyle\pm 3.73}$             & $78.6 {\scriptstyle\pm 1.46}$           \\ 
    \bottomrule
\end{tabular}
\end{table}

In summary, using APC pretraining improves performance substantially in noisy conditions, and additionally improves performance in noisy conditions. 
Our proposed DN-APC in combination with MTR achieves the best performance, with an average improvement of \SI{7.1}{\percent} in seen noise and \SI{8}{\percent} in unseen noise compared to baseline+MTR.

\section{Conclusions}
In this paper, we proposed the use of self-supervised pretraining to leverage unlabelled data for improving the robustness of a personalized VAD model in adverse conditions.
For pretraining we used the APC framework, while we also proposed a Denoising variant of APC for improved robustness.
We compared the pretrained models with a supervised baseline and tested their performance in both clean and adverse conditions with both seen and unseen noise at various SNR-levels. 

Our results show a significant improvement in robustness to background noise when using APC pretraining. 
Both APC and our proposed Denoising APC outperform the baseline, while our proposed Denoising APC achieves the best performance. 
Overall, it can be concluded that self-supervised pretraining can improve the personalized VAD performance in both clean and noisy conditions. 

\clearpage

\bibliographystyle{IEEEbib}
\bibliography{references}

\end{document}